# Power spectrum analysis of limb and disk spicule using Hinode Ca H line broadband filter

E. Tavabi

Physics Department, Payame Noor University (PNU), 19395-3697-Tehran, Iran

**Abstract**

We present observations of a solar quiet region obtained using Hinode Solar Optical telescope (SOT) in Ca II H line with broadband filter taken on November 2006. We study off-limb and on disk spicules to find a counterpart of limb spicule on the disk.

This investigation shows a strong correspondence between the limb and near limb spicules (on- disk spicules that historically were called dark or bright mottles especially when observed in Hα rather cool line) from the dynamical behavior (e.g., periodicity).

An excellent time sequence of images obtained near the equatorial region with a cadence of 8 s was selected for analysis. 1-D Fourier power spectra made at different positions on the disk and above the limb are shown. We take advantage of the so-called mad-max operator to reduce effects of overlapping and improve the visibility of these hair-like features.

A definite signature with strong power in the 3 min. (5.5 mHz.) and 5 min. (3.5 mHz) oscillations for both places exist. A full range of oscillations was found and the high frequency intensity fluctuation (greater than 10 mHz or less than 100 sec.) corresponding to the occurrence of the so- called type II spicules and, even more impressively, dominant peaks of Fourier power spectra are seen in a wide range of frequencies and for all places of "on" and "off" disk spicules, in rough agreement with what historical works were reporting regarding the disk mottles and limb spicules. Also, some statistically significant behavior, based on power spectrum computed for different positions, is discussed. The power for all kind of power spectra are decreasing with the increasing distance from the limb, except for photospheric oscillations (5 min. or p-mode), which show a dominate peak for on-disk power spectra.

### 1- Introduction

The chromosphere is the region between the temperature minimum and the corona. This layer is made almost entirely of spicules. Their interpretations were listed in review of Sterling (2000). The dynamics of fine jets is interesting and may have broad implications for mass and energy balance of the whole outer atmosphere, including the solar wind. Evaluating them continue to be a highly attracting research subjects. These structures were identified and studied at the solar limb in rather cool emission lines.

Such structures are almost dark when the background radiation from the solar disk is absorbed (Beckers 1972). Historically, some authors (Macris 1965 and de Jager 1957) identify spicules with the disk dark mottles. A traditional question is whether they are identified as disk counterparts of limb spicules or they constitute a part of limb spicules. Koutchmy and Macris (1971) applied a photographic method to make observation of mottles crossing the limb by limb-darkening correction and Georgakilas et al. (2001) use an image processing method for this purpose. From classical studies, most spicules have lifetimes of ~ 1-10 minutes (Lippincott 1957 and Tavabi et al. 2011-a) but some spicules may live longer than 45 min for macrospicules (Bohlin et al. 1975) or shorter 10 to 100 sec (De Pontieu et al. 2007).

Since the oscillations discovered by Leighton (1961), the p-mode oscillation was the subject of hundreds of studies. Periods close to 300 sec are dominants in the photosphere and the convective zone, while a period of about 200 sec is mainly visible in the chromosphere, above the temperature minimum region. Intensity fluctuations are observed mostly with 5 min period, which may show their connection to the photospheric 5 min oscillations that leakage along the inclined and rather dense dark mottles to upper layers (De Pontieu et al. 2004). Nikolsky and Platova (1971) had concluded that the limb spicules show quasi-periodic behavior with 1 min period; this period is close to the lifetime of the so-called type II spicules (Zaqarashvili and Erdelyi 2009).

Intensity oscillation in rosettes were studied by several authors, mostly the dominant 5 min periodicity (Dame et al. 1984, Baudin et al. 1996, Gupta et al. 2012).

One of the most distinct chromospheric network features are the bright points concentrated in magnetic regions. Theynare co-spatial with photospheric G-band bright points as well as the bright points seen in Ca II H-line images, where one can see higher in the solar atmosphere (Hansteen et al. 2007). Signatures of 3 and 5 mHz wave power above the photospheric bright points were found in all chromospheric layers and even up to transition region and coronal layers with less evidence.

In this study a long time series of high spatial and temporal resolution of Ca 396.8 nm H-line images is used to study intensity periodicity properties of spicules and network bright points, to clear up their dynamical characteristics and relationships and to explore their association with wave propagation phenomena.

## 2- Observations

We selected a long sequence of solar limb observations taken with the broad-band filter instrument (BFI) of the SOT of the Hinode mission (Fig 1), at the beginning of this mission (22 Nov 2006) in the resonance Ca II H emission line; the wavelength pass-band being centered at 398.86 nm with a FWHM of 0.3nm. A fixed cadence of 8s is used (with an exposure time of 0.5s) giving a spatial full resolution of the SOT/Hinode; a 0"0541 pixel size scale is used (Tsuneta et al. 2008).

The size of all images used here is 512× 512 pixels (Hinode read out only the central pixels of the larger detector to keep the high cadence in the telemetry restrictions) thus covering an area of (FOV) 56"×56" (with 0"1 pixel size it corresponds to the half of Hinode full spatial resolution) and images are centered at position x=960, y=-90 arcsec; they include more than 1000 frames. A drift with an average speed less than 0.015"/min toward the east was identified from solar limb motion.

A superior spatial processing for thread-like feature is obtained using the mad-max algorithm (Tavabi et al. 2013). The spatial filtering using the "mad-max" algorithm clearly shows fairly bright radial threads in the chromosphere as fine as the resolution limit of about 120 km, see Fig.1 and Tavabi et al. (2013) for more details about the mad-max algorithm that permits to deduce in first order approximation, what could be the individual properties of spicules.

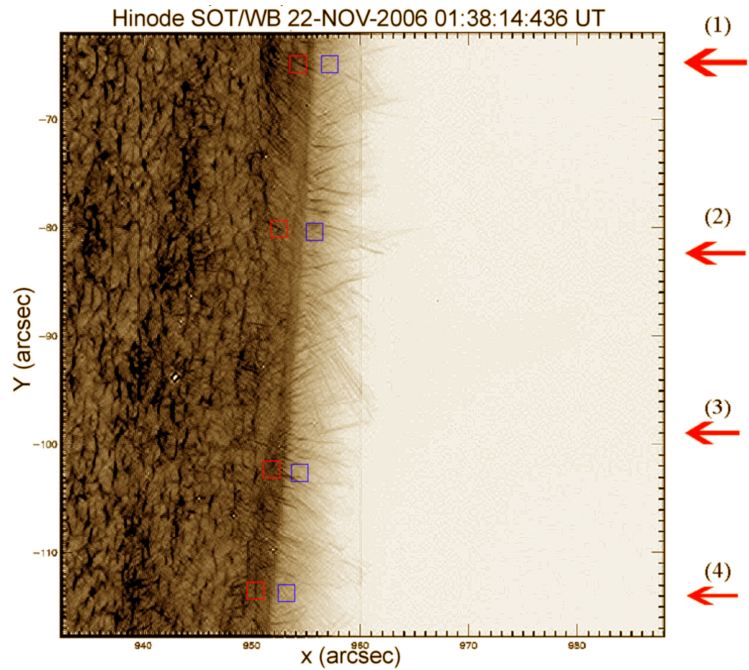

**Fig. 1** Negative image to improve the visibility of fine features after applying the mad-max operator on SOT (Hinode) filtergrams in H line of Ca II, obtained on November 22, 2006, near the equator. The positions which was selected for analysis are shown by color boxes (red for disk and blue above the limb), and the red arrows are random positions to make time-slices that was shown in figure 4.

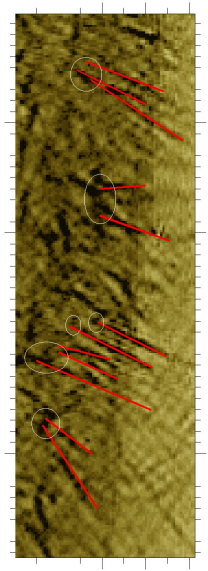

**Fig. 2** Negative sub-frame of figure 1 showing crossing spicules and bright points, red lines show the trajectory of spicules and the chromospheric network bright points (root of spicules) are marked by white rings; the scale in this image is same as in figure 1.

**3- Data analysis**

We perform a temporal Fourier transform of the filtergrams to obtain the power spectra of intensity oscillations for a selected area including disk (red boxes) and limb spicule (blue boxes in figure 1). Intensities are averaged over 10 by 10 pixels which correspond to boxes shown in figure 1. This summing process of individual regions, may the results more statistical, and reduces the noise-of the resulting spectra; finally a small spectral smoothing has been performed by dividing a long series into several segment of equal length of time, computing the power spectra for each segment, and then averaging with the use of an appropriate long window (Welch, 1967). Note that these diagrams (figures 5, 6 & 7) are not corrected for the modulation transfer function (MTF) of the instrument which attenuates the high frequency part (Tavabi et al. 2011-b). Two different power spectra were done with the same field for disk and off disk regions. As expected, the general properties and dominant peaks appear to be similar for regions that are close to the limb. The time-Y cut of four different rows shown in figure 1 (red arrows), are shown in figure 4.

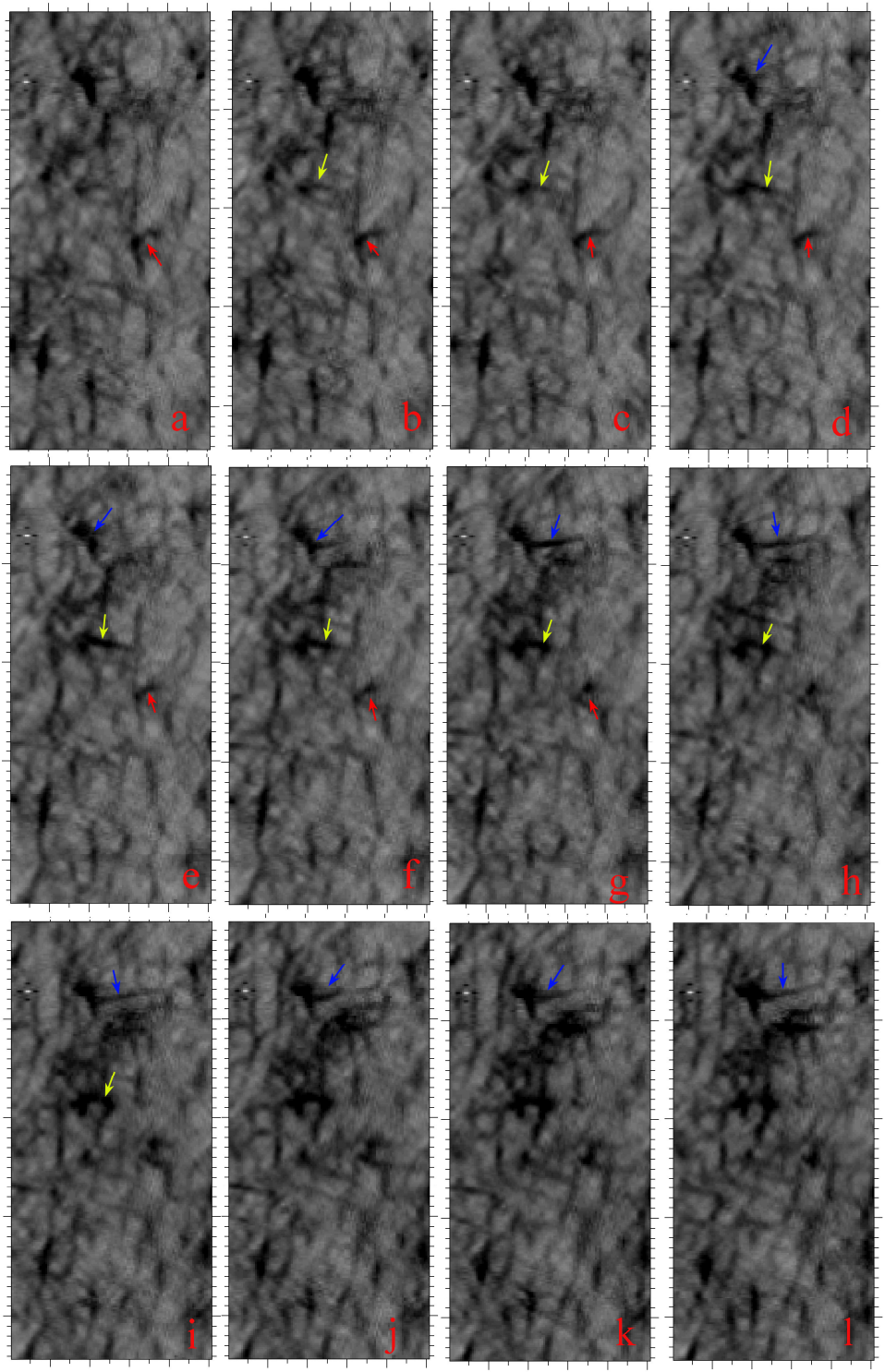

**Fig. 3** Snapshots of successive negative frames obtained with an 8 sec cadence, to show disk spicules and their connection with bright points. Almost all elongated spicules are with bright points at their end point where their roots are. Some of the most obvious examples were marked with color arrows (see online version in color).

## 4- Results

In this study, with the achievement of 120 km spatial resolution that is of order of the spicule diameter (Tavabi et al. 2011-b) and with highly processed images, we got the observational evidence for oscillation periods along spicules and at their roots simultaneously, and we find some period-brightness relationship. The power in different positions, e.g. in bright points and in bright elongated features above the limb (spicules), over rather dark regions (internetwork), shows different amplitudes in low and high frequency domains. Dunn et al. (1974) using the Sacramento Peak vacuum solar telescope in the far wings of the Hα line could not find a good correlation with bright points in rosette regions with dark mottles.

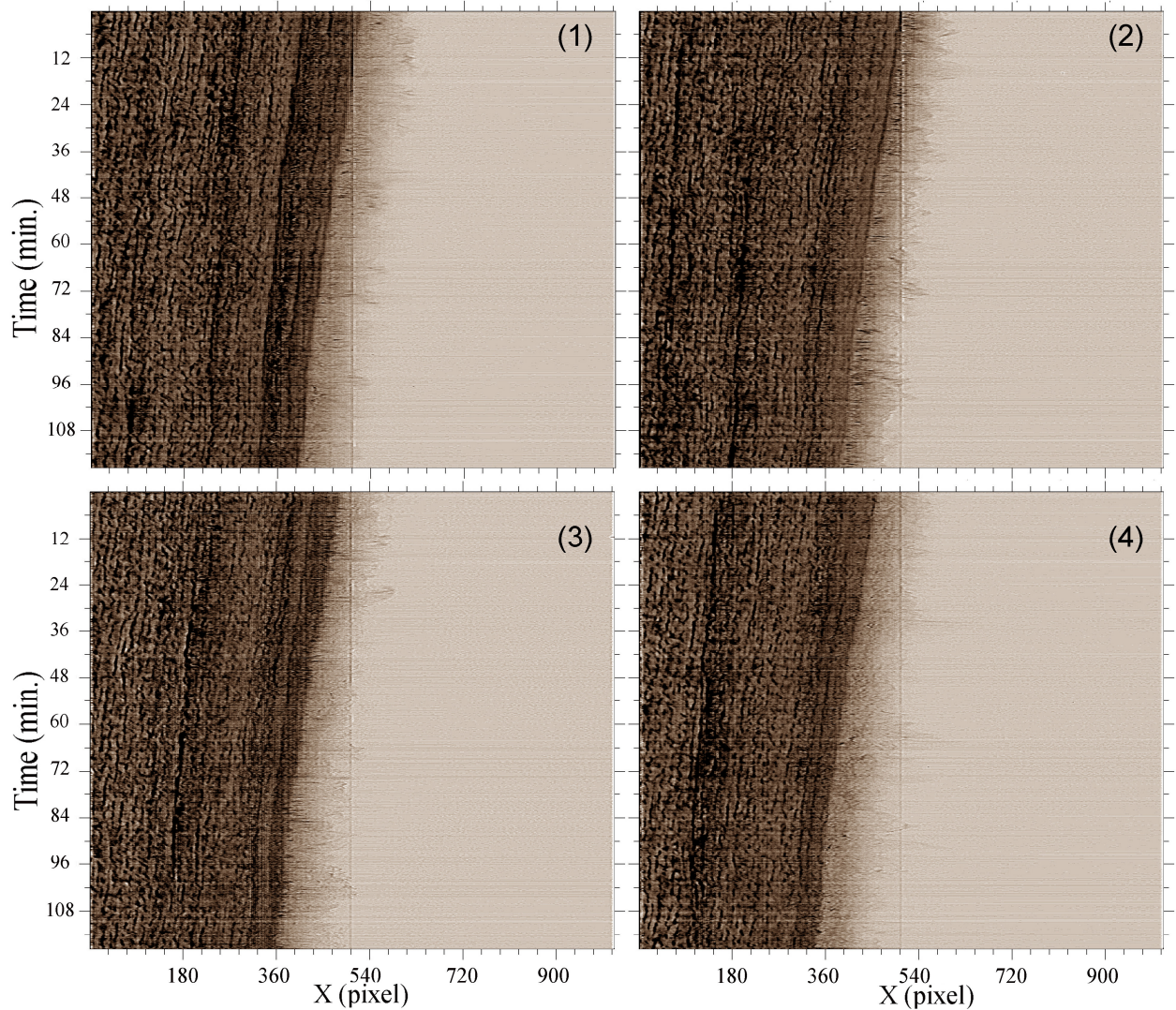

**Fig. 4** Time-Y cuts diagrams in negative for the 4 different rows shown in figure 1 by red arrows, for looking at each row in details (use online electronic version with a larger magnification please).

The fact that 5 min oscillation are of first importance and most dominant peaks is not surprising , and their recurrent behavior has been pointed out by many authors (Dame et al. 1984, Baudin et al. 1996, Ajabshirzadeh et al. 2008 and Tavabi et al. 2011-b) in the network and commonly concluded that these disturbances are either confined to the chromosphere or are excited by photospheric events (De Pontieu et al. 2004), the higher periods (low frequency<1.5 mHz) may be related to the gravity mode effect (Dame et al. 1984 and Gupta et al. 2012). If the oscillation bands have been divided as low (3-6 mHz) and high (8-14 mHz) frequency, from power spectra (figures 5 and 6) the corresponding frequency at the first dominant power peak is 3.5 mHz (5 min) whereas that at the second maximum is around of 5.5 mHz (~3 min). These are the frequency which are dominating in both regions. On disk it is more powerful than off disk. For high frequency ranges (8<f<14 mHz), the dominating peaks are around of 10 mHz (100 sec) or between 120 sec to less than 1 min for off-limb spicules. All these values are similar for the disk network (bright points). For the inter-network the 3 min oscillation was not clearly detected as a separate peak (figure 7).

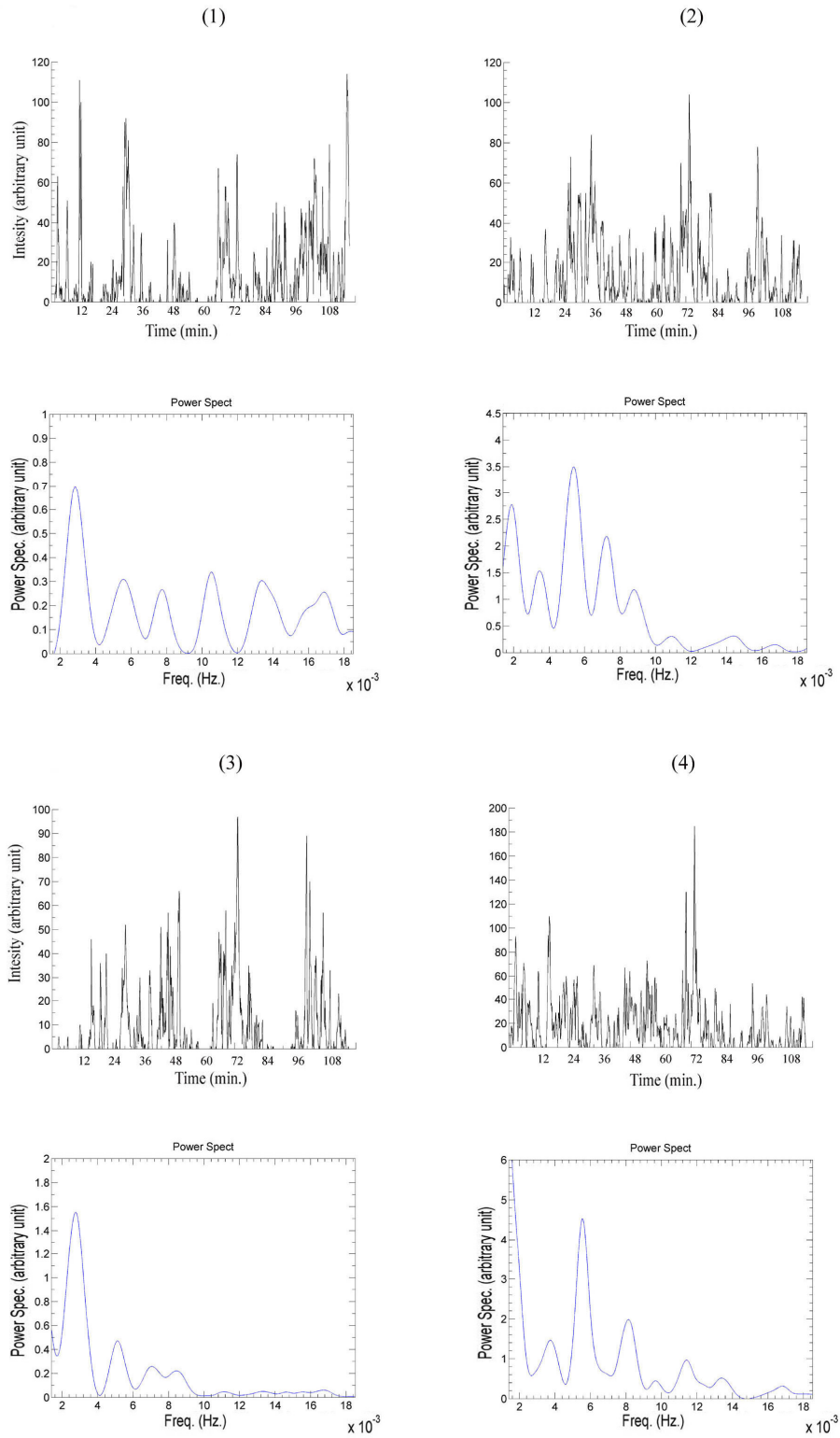

**Fig. 5** Intensity fluctuations and Fourier power spectra computed using individual summing over the selected area of a time series. The analysis is repeated the same way for all blue boxes of figure 1 above the limb.

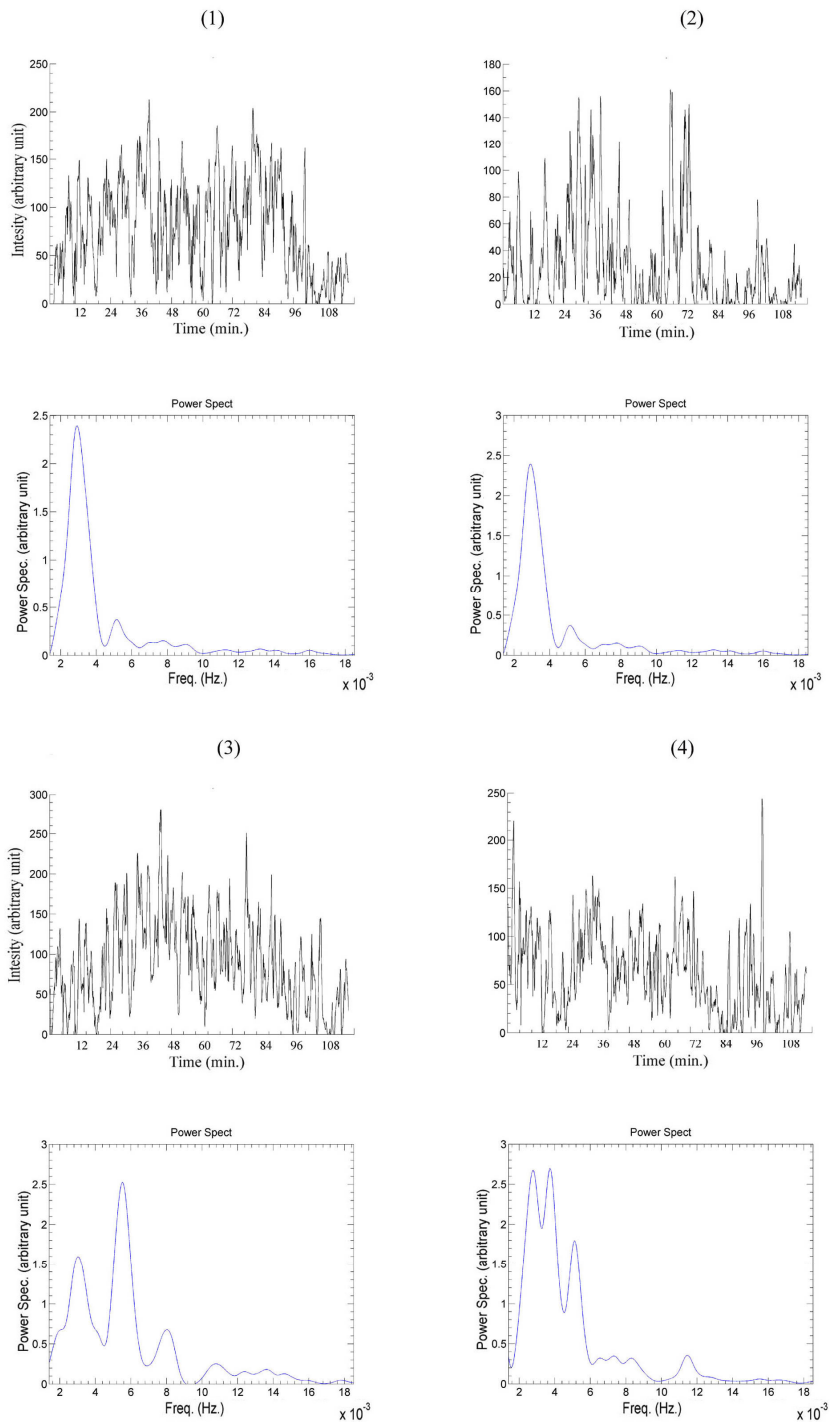

**Fig. 6** Intensity fluctuations and Fourier power spectra computed using individual summing over the selected area of a time series. The analysis is repeated the same way for all selected red boxes (see figure 1) of the disk bright features (include bright points and mottles).

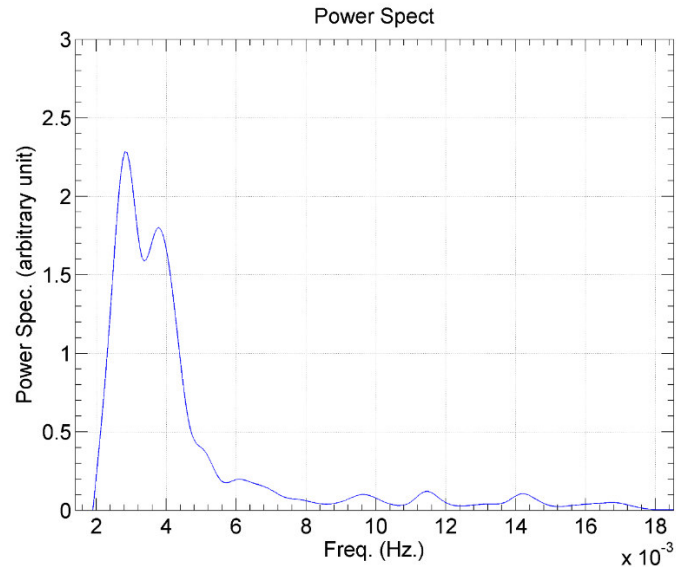

**Fig. 7** The power spectrum for a region near the center of the Sun for inter network cell without influence of mottles and network bright points intensity fluctuation effect.

**5- Discussions and Conclusions**

Since Loughhead et al. (1968), the "dark" fine mottles close to the limb in Hα line center (and wing) were considered. There crossing and connection to the bright features above the visible limb have been investigated in several works including a discussion of the relationship between spicules and chromospheric network bright points (Bhavilai 1965, Sterling et al. 2010 and etc). Thanks to Hinode/SOT unprecedented temporal (8 sec) and spatial resolution (~120 km) in chromospheric Ca II-H line and their good contrast, taking into account the limb-darkening effect, off limb spicules have been traced from outside toward the solar disk, just inside the limb, where they appear as bright emission mottles in the CaII filtergrams. In figure 2, several such cases are indicated by red lines. In figure 3, the time variations of a group of bright mottles is demonstrated. Here we should emphasis that these bright mottles are significantly different from the historic bright mottles in Hα broad line images. Several decades ago, some authors believed that the bright mottles are the lower part of elongated dark Hα mottles. Now the much higher resolution data show that these bright mottles are simply the bright background below the dark mottles (Tsiropoula et al. 2012). The morphological relationship between bright points and mottles is similar to the relation between plage and fibril (Tziotziou et al. 2004), and these features show a wide range of periodical variations, from 100 to 500 sec.
Suematsu et al. (1995) reported several bright points in Hα wing which often appear at the base of spicules (Koutchmy and Macris, 1971); more recently, Sterling et al. (2010) found that some of spicules locations on the disk are coming from Ca II brightenings with an apparent velocity of order of 10 km/s. Tavabi (2014) used a 3D time slice "column" diagrams (2D in *x* and *y* and time in *z* being the 3d dimension) for consecutive partly transparent slices put in perspective to show the rotational behavior at chromospheric rosettes, providing a wealth of information on spinning motion of order 20–30 km/s, helical wave propagation and splitting. Jess et al. (2009), reported the detection of oscillatory phenomena associated with network bright points with periodicities ranging 126 to 700 sec. Most authors (Mosher and Pope 1977 and Tanaka 1974) have applied the term spicules to the jets coming out of the magnetic element of the chromospheric network bright points. Otsuji et al. (2007), observed emerging flux using Hinode/SOT and found the size of Ca II H-line bright points is rather larger than those on the G-band filtergrams and of Stokes I co-spatial features but the magnetic concentrations are identified with corresponding bright points in both the G-band and Ca II H-line filtergrams.

A Fourier analysis based on deeply processed data, leads to an estimate of the main intensity oscillation periods of spicules and bright points on regions close to the limb. They are a valuable approach to tackle the relationship of Ca II H line bright points and spicules oscillations. As a result of this study, 3 minute oscillations are well observed out of the limb and on the disk bright regions. The relationship between these features has been looked using power spectra distribution and show similar peaks, meaning similar periodicities. Both 3 and 5 minutes oscillations are detected in bright points and mottles with dominate peaks. Fourier power spectra of the inter network regions mainly shown by dark non-magnetic regions did not show the effect of 3 min. oscillations (figure 7). In the higher frequency range 8-14 mHz, the periodicity is similar to the lifetime of the so-called type II spicules; these peaks in the power spectra may be caused by short lived spicules.

**Acknowledgment.** We thank Serge Koutchmy for useful comments, idea and remarks. We are grateful to the Hinode team for their wonderful observations. Hinode is a Japanese mission developed and launched by ISAS/JAXA, with NAOJ as domestic partner and NASA, ESA and STFC (UK) as international partners. Image processing wavelet software was provided by O. Koutchmy http://www.ann.jussieu.fr/~koutchmy/index_newE.html. This work has been supported by the Center for International Scientific Studies & Collaboration (CISSC) and by the French Embassy in Tehran and the French CNRS and IAUT (Iran).